\documentclass[doublecol]{epl2} 
\usepackage[normalem]{ulem}

\title{Out-of-equilibrium Kondo Effect in a Quantum Dot:\\Interplay of Magnetic Field and Spin Accumulation}

\shorttitle{Out-of-equilibrium Kondo effect in quantum dot with spin accumulation}

\author{Shaon Sahoo\inst{1,2} \and Adeline Cr\'epieux\inst{3} \and Mireille Lavagna\inst{1,2,4}}
\shortauthor{Shaon Sahoo \etal}

\institute{                    
  \inst{1} Universit\'e Grenoble Alpes, INAC-PHELIQS - F-38000 Grenoble, France\\
  \inst{2}CEA, INAC-PHELIQS - F-38000 Grenoble, France\\
  \inst{3} Aix Marseille Univ, Universit\'e de Toulon, CNRS, CPT - Marseille, France\\
  \inst{4}Also at: Centre National de la Recherche Scientifique - CNRS - Grenoble, France.
} 


\abstract{
We present a theoretical study of low temperature nonequilibrium transport through an interacting quantum dot in the presence of Zeeman magnetic field and current injection into one of its leads. 
By using a self-consistent renormalized equation of motion approach, we show that the injection of a spin-polarized current leads to a modulation of the Zeeman splitting of the Kondo peak in the 
differential conductance. We find that an appropriate amount of spin accumulation in the lead can restore the Kondo peak by compensating the splitting due to magnetic field. By contrast when the injected 
current is spin-unpolarized, we establish that both Zeeman-split Kondo peaks are equally shifted and the splitting remains unchanged. Our results quantitatively explain the experimental findings reported in 
KOBAYASHI T. {\it et al.}, {\it Phys. Rev. Lett.} \textbf{104}, 036804 (2010). These features could be nicely exploited for the control and manipulation of spin in nanoelectronic and spintronic devices.}

\begin{document}

\maketitle


\section{Introduction}

Progress in nano-fabrication opened the emergence of a new class of objects, semi-conductors quantum dots -QDs- in which a few electrons localized in a small spatial region are 
connected to leads through tunneling barriers. QDs are very attractive for electronic and spintronic applications due to the possibility they offer to control and manipulate the spin. 
They give the unique opportunity to observe a tunable Kondo effect at low temperature when the dot possesses an odd number of electrons and acquires a net spin S=1/2. The 
theoretical predictions of a Kondo effect in such nanostructures were made in the late 80s~\cite{ng88,glazman88}. The Kondo effect is a many-body phenomenon which 
takes place when a localized impurity with an unpaired spin is embedded in a metallic host. It arises from resonant hopping processes of the conduction electrons of the host in 
and out of a localized impurity. This resonant process leads to the screening of the spin of the localized electrons with the formation of a Kondo singlet state. The binding energy 
of this singlet state defines the Kondo temperature $T_K$. It was predicted that the Kondo effect leads to an increase of the linear conductance of the QD when temperature is 
lowered below $T_K$. This feature is the exact analog of the rise of resistivity brought by for the Kondo effect in bulk metals~\cite{hewson93} when temperature is lowered below $T_K$.
Experimentally the first observation of the Kondo effect in QDs was made in GaAs-based two-dimensional structures in the late 90s~\cite{goldhabergordon98,cronenwett98}. 

For any usefulness of nanoelectronic and spintronic devices, it is necessary to be able to control and manipulate the spin in these systems. In this perspective QDs are excellent 
candidates since their properties can be tuned in a controlled way by varying voltages. They can be placed in an out-of-equilibrium situation by applying a finite source-drain voltage 
$V_{D}$ between the two leads (by convention source voltage $V_S$ is considered as the ground potential). In the case of a single-level QD connected to normal metal leads, the 
differential conductance $g_D=dI_D/dV_D$ vs $V_{D}$ exhibits a zero-bias anomaly~\cite{vanderwiel00}. In the presence of a Zeeman magnetic field $\Delta$, the Kondo peak in 
the differential conductance is split with a value of the splitting of the order of $2\Delta /e$ as discussed in~\cite{cronenwett98,costi00,rosch03,hewson05}. The transport properties 
of the QD can also be changed by modifying the environment of the dot. A case of special interest in connection to the study presented in this Letter corresponds to QDs with ferromagnetic 
leads~\cite{zhang02,martinek03,choi04,krawiec07}. It was shown that spin-polarization in the leads results in an effective static magnetic field which splits the Kondo peak in 
$g_D$ as observed experimentally. The Kondo peak may then be restored by compensating this effective magnetic field by a Zeeman magnetic field.

\begin{figure}
\onefigure[width=7.0cm]{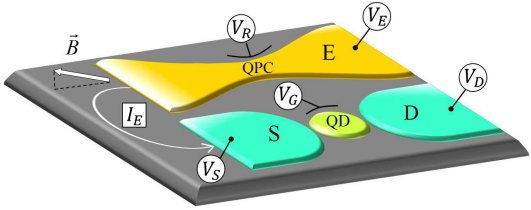}
\vspace{0.2 in}
\caption{Schematic representation of the experimental setup constituted by a QD connected to the two electrodes, source S and drain D, and a QPC responsible for the generation of 
the current $I_E$ injected from the emitter E into S. An external magnetic field is applied to the system with the plane of the device tilted by a small angle to the axis of the magnetic field.}
\label{exprt}
\end{figure}

More recently there has been a considerable effort in developing new techniques to modify the environment to achieve efficient control of the spin in QDs. The injection of a current 
in one of the leads of a QD has emerged as a very powerful way to attain this goal with the possibility to produce a spin accumulation in the lead when the current is spin-polarized
~\cite{potok02,taniyama03,katsura07,qi08,kobayashi10}. We especially refer to experimental work by Kobayashi et al.~\cite{kobayashi10} whose experimental setup is schematized 
in fig.~\ref{exprt}. The generation of the current is achieved with the aid of a quantum point contact -QPC- which is spin-polarized by applying a high parallel Zeeman magnetic field. 
The differential conductance of the QPC, $g_E=dI_E/dV_E$ vs gate voltage $V_R$ is quantized~\cite{vanwees88} at multiples of $e^2/h$ determined by the number of occupied 
subbands in the QPC. The current $I_E$ induced by the application of a bias voltage $V_E$ to the emitter E, is then magnetically focused~\cite{vanhouten89} into S along the cyclotron 
trajectory by applying a low perpendicular magnetic field. In practice in order to apply a high parallel magnetic field for Zeeman splitting in both QPC and QD, and a low 
perpendicular magnetic field for magnetic focusing, the 2DEG plane is tilted by a small angle to the axis of the applied magnetic field.

The experimentalists have shown that the low temperature transport through a Kondo QD is considerably influenced by the injection of a current into one of its leads. The 
observations~\cite{kobayashi10} show spectacular effects on the evolution of the differential conductance with $V_D$ depending on the number of open transmission channels in the 
QPC which can be controlled by $V_R$. The profile of the Zeeman-split Kondo peaks versus $V_E$ are found to have a very characteristic dependence on the nature of injected current. 
While a spin-polarized current affects the separation between the two peaks in the differential conductance, a spin-unpolarized current equally shifts both peaks. The former case thus 
offers the possibility of recovering the Kondo peak by accumulating an appropriate amount of spin in one lead to compensate the Zeeman magnetic field effect. 

On the theoretical side, the pioneering works go back to~\cite{qi08} and~\cite{lim13}. Qi et al.~\cite{qi08} examined the fate of the Kondo resonance peak in the density of states
in the presence of a spin accumulation for systems with a local impurity embedded in a metal. By using an equation of motion (EOM) approach on the single impurity Anderson model -SIAM, 
they found that the Kondo resonance is split into two peaks pinned to the spin-dependent chemical potentials. They then showed that the Kondo resonance may be restored by applying 
an external magnetic field. Since they are bulk, these systems do not offer the possibility of applying a finite bias voltage across the impurity.  Lim et al.~\cite{lim13} further considered 
the situation of quantum dots in the presence of static spin polarization of the contact and spin accumulation in the electrode as resulting from the injection of a spin-polarized current. 
By also using an equation of motion approach on the SIAM, they showed that spin polarization and spin accumulation have antagonist effects on the Kondo peak for both the spectral density 
and differential conductance. Whereas the spin-polarization of the contact is shown to introduce a splitting of the Kondo resonance, they demonstrated that the spin accumulation may compensate 
the latter splitting and restore the Kondo resonance. These two theoretical works have the merit of having highlighted the role that a spin accumulation can have on the Kondo effect. However 
we emphasize that their results have been obtained in the infinite $U$ limit of the model. Moreover in~\cite{lim13}, the truncated scheme considered within the EOM approach assumes 
$\langle f_{\sigma}^\dagger  c_{k\alpha{\sigma}} \rangle=0$ following Meir et al.~\cite{meir91, meir93}. This assumption is known to be valid in the high temperature regime when $T \geq T_K$. 
By contrast it is important to have in mind that the whole set of results obtained by Kobayashi et al. has been obtained in the low temperature regime when $T \leq T_K$ in systems where the 
Coulomb interaction is estimated to $1.5 meV$ far from the infinite $U$ limit. The results obtained therefore in the two theoretical works do not apply to the situation in which the experiments are performed.

The purpose of this Letter is precisely to fill this discrepancy and to study how the spin accumulation in one of the leads of a QD affects the transport properties of an interacting quantum 
dot in the low temperature and finite $U$ regime. To do this we choose to carry out our theoretical study in conditions as close as possible to those in which the experiments were carried out. 
Our calculations based on the single impurity Anderson model at finite $U$ are performed by using the self-consistent renormalized equation of motion approach following the scheme developed 
in \cite{lavagna15,lavagna} in nonequilibrium situation. The decoupling scheme used to truncate the set of EOM considers the mixed decoupling parameter $\langle f_{\sigma}^\dagger  c_{k\alpha{\sigma}} \rangle$ 
in addition to the usual decoupling parameters $\langle c_{k'\alpha{\bar{\sigma}}}^\dagger  c_{k\alpha{\bar{\sigma}}} \rangle$ and $\langle n_ {\bar{\sigma}} \rangle$. This additional decoupling parameter 
plays a key role in the description of the strong coupling regime reached at low temperature. It can be viewed as a pseudo-order parameter which gets finite in the strong coupling regime, reminding of 
the slave-boson introduced in auxiliary-field approaches. Moreover the scheme includes two major improvements related to the renormalization of intermediate state inverse lifetimes and the 
renormalization of dot energy level, defining the self-consistent renormalized EOM approach. The renormalization of the intermediate state inverse lifetimes allows to cure the long-standing problem 
about the presence of a spurious peak in the density of states. This unphysical peak just compensates the actual Kondo resonance peak at the particle-hole symmetric point $\varepsilon_{\sigma}=-U/2$, 
therefore prohibiting one from studying the Kondo physics at this point. This serious drawback of the standard EOM approach is avoided in the self-consistent renormalized approach used in this work. 
Let us note that the particle-hole symmetric limit corresponds precisely to the situation in which the experimentalists have conducted their experiments where the system is placed at the middle of the Kondo 
conductance valley. Our calculations show that the splitting of the Kondo peak in the differential conductance is modulated by the shift of the chemical potentials introduced by spin injection. The results for the 
differential conductance vs $V_D$ and $V_E$ are found to be in quantitative agreement with the experimental results. We analyze them in detail by extracting the Kondo peak parameters and comparing 
them with the parameters extracted from experiments.


\section{Model}

The QD is modeled by the single impurity Anderson model
\begin{eqnarray}
\label{andersonhamiltonian}
\nonumber
H&=&\sum_{k,\alpha \in (S,D) ,\sigma}{\varepsilon_{k\alpha\sigma}c_{k\alpha\sigma}^{\dag}c_{k\alpha\sigma}}+\sum_{\sigma}{\varepsilon_{\sigma}f_{\sigma}^{\dag}f_{\sigma}}
+Un_{\uparrow}n_{\downarrow}\\
&+&\sum_{k,\alpha \in (S,D) ,\sigma}{(t_{\alpha\sigma}c_{k\alpha\sigma}^{\dag}f_{\sigma}+h.c.)}~,
\end{eqnarray}
where $c_{k\alpha\sigma}^{\dag}$ ($c_{k\alpha\sigma}$) creates (annihilates) an electron with momentum $k$, spin $\sigma$ ($\sigma=\pm1$) and energy $\varepsilon_{k
\alpha\sigma}$ in the $\alpha$ lead.  $f_{\sigma}^{\dag}$ ($f_{\sigma}$) creates (annihilates) an electron with spin $\sigma$ and energy $\varepsilon_{\sigma}=
\varepsilon_{0}-\sigma\Delta/2$ in the dot where $\Delta=|g^{*} \mu_B B|$ is the absolute value of the Zeeman splitting with $g^{*}$  the g-factor in GaAs~\cite{note1} and $
\mu_B$ the Bohr magneton. $U$ is the on-site Coulomb interaction in the dot. $n_{\sigma}=f_{\sigma}^{\dag}f_{\sigma}$ and $t_{\alpha\sigma}$ is the transfer matrix element 
between states, assumed to be k-independent.

In the steady state the current through the dot for spin $\sigma$ is given by~\cite{meirwingreen92}, 
\begin{eqnarray}
\label{meirwingreen}
\nonumber
I_{D\sigma}&=&\frac{2e}{\hbar} \int_{-W}^{+W} {d\omega \widetilde{\Gamma}_{\sigma}(\omega)}\\
&\times&[n_{F}(\omega-\mu_{L\sigma})-n_{F}(\omega-\mu_{R\sigma})] A_{\sigma}(\omega),
\end{eqnarray}
where $\widetilde{\Gamma}_{\sigma}(\omega)=\frac{\Gamma_{L\sigma} (\omega)\Gamma_{R\sigma} (\omega)}{\Gamma_{L\sigma} (\omega)+\Gamma_{R\sigma} 
(\omega)}$ with the tunnel coupling constants given by $\Gamma_{\alpha\sigma} (\omega)=\pi { | t_{\alpha\sigma}| }^{2} 
\rho_{\alpha\sigma}^{0}(\omega)$. $\rho_{\alpha\sigma}^{0}(\omega)$ is the density of states in the $\alpha$ lead for spin $\sigma$ and $W$ is the half-bandwidth.
$A_{\sigma}(\omega)=-\frac{1}{\pi} \mathrm{Im} G_{\sigma}^{r} (\omega)$ and $G_{\sigma}^{r} (\omega)$ are respectively the spectral density and retarded Green 
function in the dot. $n_F(\omega-\mu_{\alpha\sigma})=[exp[(\omega-\mu_{\alpha\sigma})/k_BT)]+1]^{-1} $ is the Fermi-Dirac distribution function in the $\alpha$ lead 
with chemical potential $\mu_{\alpha\sigma}$. $\mu_{D\sigma}=\mu_0-eV_D$ for both spin $\sigma$ where $\mu_0$ is the chemical potential at equilibrium. When 
the lead S is exposed to a current injection, the chemical potentials in S are selectively shifted depending on the value of $g_E$. When the QPC is tuned in the middle 
of the 0th plateau, $g_E=0$, no current goes through the QPC and $\mu_{S\uparrow}=\mu_{S\downarrow}=\mu_0$. When the QPC is tuned in the middle of the 1st plateau, 
$g_E=e^2/h$, a spin-polarized current with only spin-up electrons is injected into S and $\mu_{S\uparrow}=\mu_0-eV_E$ whereas $\mu_{S\downarrow}=\mu_0$. When the 
QPC is tuned in the middle of the 2nd plateau, $g_E=2e^2/h$, the current is spin-unpolarized and $\mu_{S\uparrow}=\mu_{S\downarrow}=\mu_0-eV_E$.


\section{Equation of motion approach}

The spectral density, $A_{\sigma}(\omega)$, appearing in eq.~(\ref{meirwingreen}) can be derived from $G_{\sigma}^{r} (\omega)$ which we evaluate using the EOM 
approach. Extensively used in the past to study bulk metals~\cite{appelbaum69,lacroix81} and quantum impurities in equilibrium~\cite{meir91}, the EOM approach has 
been more recently extended to nonequilibrium~\cite{meir93,entinwohlman05,monreal05,kashcheyevs06,swirkowicz06,qizhu09,vanroermund10,lim13,lavagna15}. We 
use here the self-consistent renormalized EOM approach as developed in~\cite{lavagna15,lavagna}. In this approach the set of equations of motion of Green functions are truncated 
at the third level of the hierarchy by performing a decoupling in terms of all possible two-operator correlation functions with equal-spin, $\langle f_{\bar{\sigma}}^\dagger  c_{k\alpha{\bar{\sigma}}} \rangle$, 
$\langle c_{k'\alpha{\bar{\sigma}}}^\dagger c_{k\alpha{\bar{\sigma}}} \rangle$ and $\langle n_ {\bar{\sigma}} \rangle$ where $\bar{\sigma}=-\sigma$. We point out the importance of
considering the mixed decoupling parameter $\langle f_{\bar{\sigma}}^\dagger  c_{k\alpha{\bar{\sigma}}} \rangle$ -undeservedly neglected most often in the literature- to properly 
describe the strong coupling regime at low temperature. This leads to the following result~\cite{lavagna15}  
\begin{eqnarray}
\label{fgreenfunction1}
\nonumber
G_{\sigma}^{r}(\omega)&=&\frac{1-\langle n_{\bar{\sigma}} \rangle}{\omega-\varepsilon_\sigma-\Sigma_{\sigma}^0(\omega)-\Pi_{\sigma}^{(1)}(\omega)}\\
&+&\frac{\langle n_{\bar{\sigma}} \rangle} {\omega-\varepsilon_\sigma-U-\Sigma_{\sigma}^0(\omega)-\Pi_{\sigma}^{(2)}(\omega)},
\end{eqnarray}
where $\Sigma_{\sigma}^0(\omega)=-i \sigma (\omega)$ and $\Gamma_\sigma(\omega)=\sum_{\alpha=S,D} \Gamma_{\alpha \sigma }(\omega)$. In the wide band limit, $\Sigma_{\sigma}^0(\omega)$ is 
independent of $\omega$ taking the value $-i \Gamma_\sigma$. $\Pi_{\sigma}^{(1)}(\omega)$ and $\Pi_{\sigma}^{(2)}(\omega)$ are defined as
\begin{eqnarray}
\label{Pi1}
&\Pi_{\sigma} ^{(1)}(\omega)&=-U \frac{\Sigma_{\sigma}^{(1)}(\omega)-(\omega-{\varepsilon}_\sigma)\Sigma_{\sigma }^{(4)}(\omega)}{\omega-\varepsilon_\sigma-U
-\Sigma_{\sigma}^{(3)}(\omega)+U\Sigma_{\sigma}^{(4)}(\omega)}~,\\
\label{Pi2}
&\Pi_{\sigma}^{(2)}(\omega)&=U \frac{\Sigma_{\sigma}^{(2)}(\omega)+(\omega-{\varepsilon}_\sigma-U)\Sigma_{\sigma}^{(4)}(\omega)}{\omega-\varepsilon_\sigma
-\Sigma_{\sigma}^{(3)}(\omega)+U\Sigma_{\sigma}^{(4)}(\omega)}~,
\end{eqnarray}
where 
\begin{eqnarray}
\label{Sigmai}
\nonumber
\Sigma_{\sigma}^{(i)}(\omega)&=& \sum_{k,\alpha} { |t_{\alpha\bar{\sigma} }|^2 \Big\lbrack} \frac{A_{k\alpha\sigma}^{(i)}}{\omega+ \widetilde{\varepsilon}_{\bar{\sigma}}-\widetilde
{\varepsilon}_\sigma-{\varepsilon}_{k \alpha\bar{\sigma}}+i\widetilde{\gamma}_\sigma}\\
&+&\frac{A_{k\alpha\sigma}^{\prime (i)}}{\omega+\widetilde{\varepsilon}_{k \alpha\bar{\sigma}}-\widetilde{\varepsilon}_\sigma- \widetilde{\varepsilon}_{\bar{\sigma}}-U+i\widetilde
{\gamma}_D} { \Big\rbrack}~,
\end{eqnarray}
with $A_{k\alpha \sigma}^{(1)}= \sum_{k'} \langle c_{k'\alpha{\bar{\sigma}}}^\dagger  c_{k\alpha {\bar{\sigma}}} \rangle$, $A_{k\alpha \sigma}^{(2)}=  1- \sum_{k'\alpha } \langle c_
{k'\alpha{\bar{\sigma}}}^\dagger  c_{k\alpha{\bar{\sigma}}} \rangle $, $A_{k\alpha \sigma}^{(3)}=1$, and $A_{k\alpha \sigma}^{(4)}=  \langle f_{\bar{\sigma}}^\dagger  c_{k\alpha{\bar
{\sigma}}} \rangle / t_{\alpha\bar{\sigma}}$. $A_{k\alpha \sigma}^{\prime (i)}=(A_{k\alpha \sigma}^{(i)})^{*}$ for $i=1,2,3$ and $A_{k\alpha \sigma}^{\prime (4)}=-(A_{k\alpha \sigma}^{(4)})
^{*}$. 

Expression ~(\ref{fgreenfunction1}) for $G_{\sigma}^{r}(\omega)$ is exact both in the noninteracting limit ($U=0$) and in the isolated-site limit ($t_{\alpha\sigma}=0$). The expression exhibits
two poles at $\varepsilon_\sigma$ and $(\varepsilon_\sigma +U)$ corresponding to the isolated-site limit, weighted by the factors $(1-\langle n_{\bar{\sigma}} \rangle)$ and ${\langle n_{\bar{\sigma}} \rangle}$ 
respectively. $\Sigma_{\sigma}^0(\omega)$ is the ordinary self-energy due to electron tunneling between the dot and the leads, whereas $\Pi_{\sigma}^{(1)}(\omega)$ and $\Pi_{\sigma}^{(2)}(\omega)$
are the self-energy contributions due to interactions. Expression ~(\ref{fgreenfunction1}) constitutes an extension of Lacroix'~\cite{lacroix81} and Meir et al.'s~\cite{meir91} results.
At equilibrium and in the infinite $U$ limit, the expression gives back the results of ~\cite{lacroix81}. When $\langle f_{\bar{\sigma}}^\dagger  c_{k\alpha{\bar{\sigma}}} \rangle=0$
(and hence $\Sigma_{\sigma}^{(4)}(\omega)=0$), the results of ~\cite{meir91} are recovered, corresponding to the high temperature limit. The consideration of this extra-parameter $\langle f_{\bar{\sigma}}^
\dagger  c_{k\alpha{\bar{\sigma}}} \rangle$ is crucial to describe the low-temperature limit. It ensures the unitary condition for $G_{\sigma}^{r}(\omega)$ at the Fermi level to be fulfilled at zero temperature~\cite
{lacroix81,lavagna15,lavagna}. The decoupling parameters $\langle f_{\bar{\sigma}}^\dagger  c_{k\alpha{\bar{\sigma}}} \rangle$, $\langle c_{k'\alpha{\bar{\sigma}}}^\dagger  c_{k\alpha{\bar{\sigma}}} \rangle$, and 
$\langle n_{\bar{\sigma}} \rangle$ are then determined by the self-consistent equations established both at and out-of-equilibrium~\cite{lavagna15} provided that the system is in a steady state. As a result the self-
energies $\Sigma_{\sigma}^{(i)}(\omega)$ are expressed in terms of  $G_{\sigma}^{r}(\omega)$. The Green function $G_{\sigma}^{r}(\omega)$ can then be self-consistently calculated from 
eq.~(\ref{fgreenfunction1}). 

We consider two important improvements related to the renormalization of both intermediate state inverse lifetimes and dot energy level. 
These two improvements define the self-consistent renormalized EOM approach where propagators and vertices of the corresponding skeleton Feynman diagrams are dressed by 
self-energy and vertex corrections respectively. In the standard EOM approach, $\widetilde{\varepsilon}_\sigma$ is the bare energy level $\varepsilon_\sigma$ in the dot,  and $\widetilde{\gamma}_\sigma$ 
and $\widetilde{\gamma}_{D}$ are both an infinitesimal positive ($\gamma_\sigma=\gamma_{D}=+i\delta$).  They are renormalized in the self-consistent renormalized
EOM approach. On the one hand, $\widetilde {\varepsilon}_\sigma$ is renormalized by self-energy corrections according to: $\widetilde {\varepsilon}_\sigma=\varepsilon_\sigma+\Re\Sigma_{\sigma}^{(1)}(\omega=
\widetilde {\varepsilon}_\sigma)$. At the particle-hole symmetric point the renormalization effect on $\widetilde{\varepsilon}_\sigma$ is zero and $\widetilde{\varepsilon}_\sigma=\varepsilon_\sigma$.
On the other hand $\widetilde{\gamma}_\sigma$ and $\widetilde{\gamma}_{D}$ are replaced by the inverse lifetimes of intermediate states. They are determined by
using the generalized Fermi golden rules up to the forth order in $t_{\alpha\sigma}$ following ~\cite{lavagna15,lavagna}, extending to finite $U$ the argument used in ~\cite{meir93} for the infinite-$U$ limit.
The renormalization of $\widetilde{\gamma}_{D}$ proves to be extremely important  to cure the long-standing problem about the presence of a spurious peak in the density 
of states. This unphysical peak, which compensates the actual Kondo resonance peak, is the reason behind the failure of the standard EOM approaches.
This drawback is avoided in the self-consistent renormalized EOM approach used in this work. By using Eqs.~(\ref{meirwingreen}-\ref{Sigmai}), we have all the ingredients to derive the total current $I_D$ and the differential conductance $g_D=dI_D/dV_D$.


\section{Choice of parameters and Kondo temperature}

Except for $U$, the values of all the parameters inserted in our model are adopted from the estimations made in~\cite{kobayashi10}. Hence the electronic temperature is taken 
as $T=100$ mK, $\Delta = 130$ $\mu$eV and $\Gamma_{\alpha \sigma} = 0.25$ meV. As far as $U$ is concerned, we choose to take a slightly larger value $U = 3$ meV 
instead of $U = 1.5$ meV considered in~\cite{kobayashi10} to ensure that the system is in the Kondo regime on the following criterion: $2\Gamma_{\alpha \sigma}\ll U/2$. 
Besides we consider the system at the particle-hole symmetric point with $\varepsilon_{0} =- U/2$ in agreement with the experimental situation.

The Kondo temperature, $T_K$, of the QD is estimated from the linear conductance vs temperature plotted at equilibrium (for $B=V_E=0$). $T_K$ is the temperature at which the 
linear conductance falls down to half of its maximum value. We get: $T_K=0.5$ K. Upper bounds to $T_K$ can be found in various nonequilibrium situations. For example, an upper 
bound to $T_K$ is estimated from the value of the FWHM of the Kondo peak in $g_D$ vs $V_D$ plot. We perform calculations at $T=100$ mK (for 
$B=V_E=0$), and get $T_K<0.7$ K. Finally the value of $T_K$ estimated from Haldane's formula~\cite{haldane78} is $0.9$ K. These values are consistent with the upper bound 
$0.7$ K estimated in experiment~\cite{kobayashi10} even though we have taken a slightly different value of $U$. Let us also mention that all our numerical calculations are performed 
at $100$ mK, well below the estimated $T_K$.

\begin{figure} 
\onefigure[width=7cm,height=12cm]{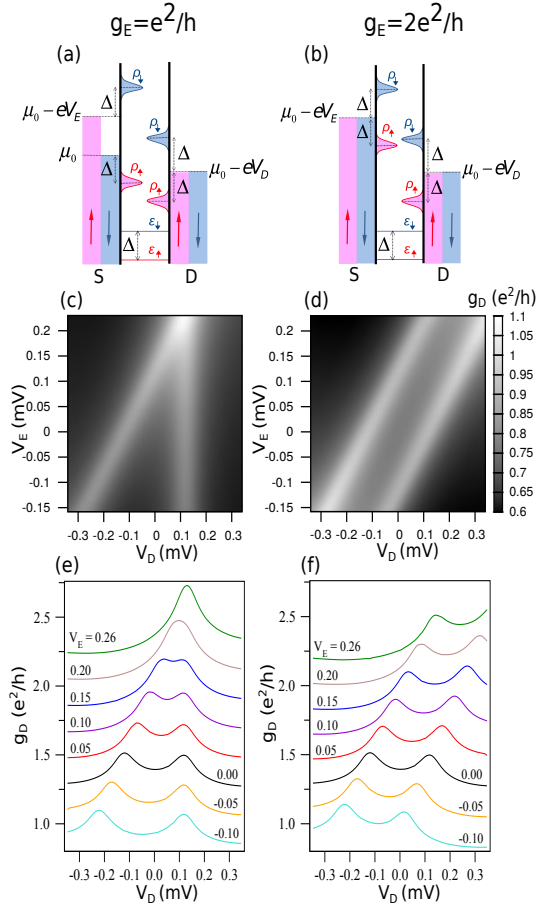}
\caption{(a)-(b) Schematic representation of the energy level diagram in the QD at $g_E = e^2/h$ and $g_E = 2e^2/h$ for finite $V_D$, $V_E$ and $\Delta$. $\rho_{\uparrow}$ 
($\rho_{\downarrow}$) represent the Kondo peaks in $A_{\uparrow}(\omega)$ ($A_{\downarrow}(\omega)$). (c)-(d) Results for the differential conductance $g_D$ in gray-scale 
representation in the plane ($V_D$,$V_E$) at $g_E = e^2/h$ and $g_E = 2e^2/h$. (e)-(f) Results for the differential conductance $g_D$ vs $V_D$ at $g_E = e^2/h$ and 
$g_E = 2e^2/h$ for $V_E$ ranging from -0.10 mV (bottom) to 0.26 mV (top). The curves are vertically offset by 0.2$e^2/h$ for clarity. The results are obtained for the symmetric 
Anderson model at $T$ = 100 mK with $U$ = 3 meV, $\Gamma_{\alpha \sigma}$ = 0.25 meV and $\Delta$ = 0.13 meV.}
\label{dfcg4}
\end{figure}


\section{Results and discussion}

Our numerical results for the differential conductance $g_D$ are represented in both gray-scale representation in the plane ($V_D$,$V_E$) in figs.~\ref{dfcg4}(c)-\ref{dfcg4}(d), and in 
$g_D$ vs $V_D$ plots in figs.~\ref{dfcg4}(e)-\ref{dfcg4}(f) at $g_E = e^2/h$ and $g_E = 2e^2/h$ respectively. We do not show the result for $g_E = 0$ since $g_D$ vs $V_D$ plot is simply 
the one obtained for $g_E \ne 0$ at $V_E = 0$. As can be seen from figs.~\ref{dfcg4}(e)-\ref{dfcg4}(f), generally $g_D$ vs. $V_D$ has two peaks. The variations of the positions of the peaks 
with $V_E$ depend on the spin-polarization state of the injected current. At $V_E=0$ the Kondo peaks occur at $V_D = \pm \Delta/e$ as expected. When the injected current is spin-polarized 
(by tuning the QPC at $g_E= e^2/h$), the position of the upper-$V_D$ peak does not vary with $V_E$ whereas that of the lower-$V_D$ peak is linearly shifted by $V_E$. The separation 
between the two peaks decreases with increasing positive $V_E$ until vanishing at a critical value of $V_E$. When the injected current is spin-unpolarized (by tuning the QPC at 
$g_E= 2e^2/h$), the positions of both peaks are equally shifted by $V_E$. 

With the aim of understanding the physical mechanisms behind these results, we illustrate in figs.~\ref{dfcg4}(a)-\ref{dfcg4}(b) the schematic representation of the energy 
level diagram in the QD at $g_E = e^2/h$ and $g_E = ­2e^2/h$ respectively, for finite $V_D$, $V_E$ and $\Delta$. From eq.~(\ref{fgreenfunction1}) it is 
easy to see that the spectral density $A_{\sigma}(\omega)$ exhibits two Kondo peaks at about $(\mu_{\alpha\bar{\sigma}}+\varepsilon_{\sigma}-\varepsilon_{\bar{\sigma}})=
\mu_{\alpha\bar{\sigma}}-\sigma \Delta$ for each $\alpha$~\cite{note2}. Following eq.~(\ref{meirwingreen}), $g_D$ vs $V_D$ exhibits a peak whenever one of the chemical 
potentials for a given spin gets aligned with a Kondo DOS peak for the same spin. This occurs when $\mu_{\beta\sigma}=(\mu_{\alpha\bar{\sigma}}+\varepsilon_{\sigma}-\varepsilon_{\bar{\sigma}})$, 
leading to the analytic prediction for the positions of the Kondo peaks. At $g_E = 0$ the Zeeman-split Kondo peaks is found to occur at $V_D = \pm (\varepsilon_{\sigma}-\varepsilon_{\bar{\sigma}})/e= \pm \Delta/e$. 
The splitting is equal to $2\Delta/e$. At $g_E= e^2/h$, the two Kondo peaks are found to be located at $V_D=-\Delta /e +V_E$ and $V_D=\Delta /e$. The separation between these 
two peaks is $(2 \Delta /e - V_E)$, which decreases with increasing $V_E$. When $V_E=2\Delta /e$, the Zeeman splitting of the Kondo peak is exactly compensated by spin 
accumulation in the lead produced by the injection of a spin-polarized current. At this compensation point, the two peaks merge into a single peak and the Kondo peak is restored. 
This manifestation can be viewed as the fingerprint of the formation of the Kondo spin-singlet state at low temperature. At $g_E= 2e^2/h$, the analytic predictions for the positions of 
the two Kondo peaks are $V_D= -\Delta /e + V_E$ and $V_D= \Delta /e + V_E$. The separation between peaks is $2\Delta /e$, independent of $V_E$.

\begin{figure}
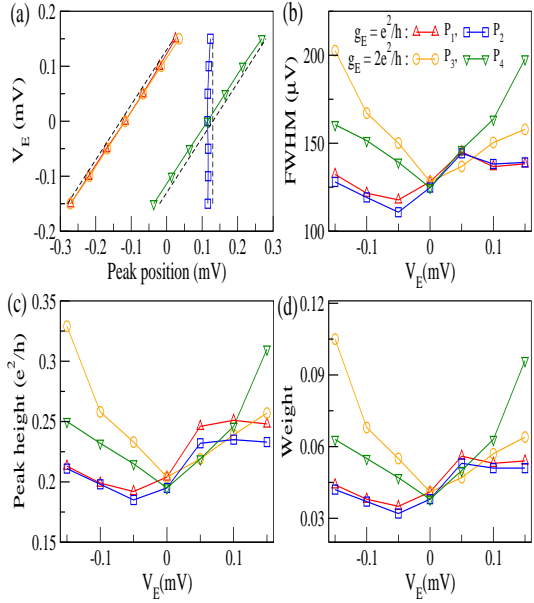

\onefigure[width=7cm,height=8cm]{figure3.eps}
\caption{Kondo peak parameters extracted from results for $g_D$ vs $V_D$. (a) Peak positions. The extracted 
peak positions are represented in solid lines whereas our analytical predictions $V_D = V_E \pm \Delta/e$ and $V_D = \Delta/e$ are represented in broken lines. 
(b) FWHMs. (c) Peak heights. (d) Weight factors.}
\label{4fig}
\end{figure}

In order to extract the peak parameters from our numerical results, we fit the curves in figs.~\ref{dfcg4}(e)-\ref{dfcg4}(f) by a double-Lorentzian function with a quadratic 
background according to: $f(x) = a + b x + c x^2 + z_1\left[\frac{1}{\pi}\frac{w_1/2}{(x-x_1)^2+(w_1/2)^2}\right]+ z_2\left[\frac{1}{\pi}\frac{w_2/2}{(x-x_2)^2+
(w_2/2)^2}\right]$. The quadratic background is necessary to account for the contributions of the two broad charge peaks in the DOS. We take two different weight factors 
$z_1$ and $z_2$ to account for the asymmetry in the spectral density arising mainly from charge accumulation in S when $V_E \ne 0$. The extracted values for peak 
positions $x_i$ (i=1,2), FWHMs $w_i$, heights $2z_i/(\pi w_i)$ and weight factors $z_i$ are reported in fig.~\ref{4fig}. It is worth noticing that the parameter extraction is 
possible only up to $V_E = 0.15$ mV. Beyond this value the two peaks are too close and can no longer be resolved. As expected, $w_i$, heights and weight factors of the 
two peaks coincide at $V_E=0$ in both cases. In fig.~\ref{4fig}, $P_1$ and $P_2$ correspond respectively to the lower-$V_D$ and upper-$V_D$ peaks at $g_E=e^2/h$ whereas 
$P_3$ and $P_4$ are the equivalent peaks at $g_E=2 e^2/h$. As can be seen in fig.~\ref{4fig}(a), our numerical results for the peak positions (in solid lines) are in 
excellent agreement with our analytical predictions of $V_D = V_E \pm \Delta/e$ and $V_D = \Delta/e$ (in broken lines). The extracted FWHMs vs $V_E$ for the different peaks are 
reported in fig.~\ref{4fig}(b). The values of the FWHM give us some useful information about the degree of decoherence in the Kondo resonance. The higher $V_D$, $\Delta$ or 
$(\mu_{\alpha\sigma}-\mu_0)$, the higher the FWHM. As can be seen from fig.~\ref{4fig}(b), $w_i$ for both $P_1$ and $P_2$ saturate at large positive values of $V_E$ when 
the system gets closer to the compensation point where the Kondo peak is restored. From the same figure, one can see that $w_i$ for $P_3$ and $P_4$ do not show any evidence of saturation at large 
values of $V_E$ as expected. Finally the extracted peak heights and weight factors vs $V_E$ are reported in figs.~\ref{4fig}(c)-\ref{4fig}(d). While the peak height results from the two 
antagonistic effects brought by $z_i$ and $1/w_i$ contributions respectively, the results show that the dominant contribution is provided by $z_i$. 

The orders of magnitude of the various peak parameters and their overall evolution as a function of $V_E$ are in good agreement with the experimental data~\cite{kobayashi10} 
although the value that we adopted for $U$ is slightly different from the experimental estimation. However we would like to point out that unlike what we find in our calculations, the 
experimental results show a deflection of the $P_2$ line from $V_D=\Delta/e$ in the vicinity of the Kondo compensation point along with large and sudden fluctuations of the FWHMs 
for both $P_1$ and $P_2$ in this range. One of the reasons for this behavior as suggested in Ref.~\cite{kobayashi10} is that the system is in a highly nonequilibrium situation 
when $g_E = e^2/h$ and hence the fermion states below $\mu_{S\uparrow}$ along the cyclotron trajectory from E to S, are not fully occupied at zero temperature~\cite
{pothier97,defranceschi02}. Ihis would result in a double-step instead of the single-step Fermi-Dirac distribution function considered in our calculations. It would be interesting in the 
future to investigate consequences of this situation.

											 
\section{Conclusion}

In summary, we have studied the combined effects of Zeeman magnetic field and current injection into one lead on the nonlinear conductance of a QD in the low temperature regime. 
When the injected current is spin-polarized, the Zeeman splitting of the Kondo peak in the differential conductance is found to be compensated by an appropriate amount of spin accumulation in 
the lead and the Kondo peak is restored in good agreement with experimental data~\cite{kobayashi10}. Our results in this Letter show that the injection of a current in one lead of a QD offers a 
new and promising route to controlling and manipulating spin in nanoelectronic devices. Present work opens the possibility of studying other important situations such as separate spin accumulations 
in both leads with or without the presence of magnetic field. In the absence of magnetic field, we predict that the Kondo peak is restored when the two leads have an equal amount of spin accumulation 
with opposite spin orientation.

\acknowledgments
We would like to thank H.~Baranger for valuable discussions. For financial support, the authors acknowledge the Indo-French Centre for the Promotion of 
Advanced Research (IFCPAR) under Research Project No.4704-02 and the Nanosciences Foundation of Grenoble under Contract CORTRANO.



\begin{thebibliography}{0}
\bibitem{note1}
{Note the minus sign in front of $\sigma\Delta/2$ in the expression of $\varepsilon_{\sigma}$ due to the fact that $g^{*}=-0.44$ in GaAs}.
\bibitem{note2}
{The fact that the Kondo peak in the density of states $A_{\sigma}(\omega)$ occurs at $(\mu_{\alpha\bar{\sigma}}-\sigma \Delta)$ reflects the formation of the Kondo spin-singlet state at low temperature}.
\end{thebibliography}
\end{document}